\documentclass[5p,twocolumn,times]{elsarticle}

\usepackage{graphicx}
\usepackage{amsmath}   
\usepackage{subfig}

\begin{document}

\begin{frontmatter}

\title{Fiber Bragg Grating (FBG) sensors as flatness and mechanical stretching sensors}



\author[add18]{D.~Abbaneo}
\author[add18]{M.~Abbas}
\author[add2]{M.~Abbrescia}
\author[add9]{A.A.~Abdelalim}
\author[add14]{M.~Abi Akl}
\author[add8]{O.~Aboamer}
\author[add16]{D.~Acosta}
\author[add20]{A.~Ahmad}
\author[add9]{W.~Ahmed}
\author[add20]{W.~Ahmed}
% ...

\cortext[cor]{Corresponding author}

% ...
\address[add2]{INFN Bari and University of Bari, Bari, Italy}
% ...
\address[add8]{Academy of Scientific Research and Technology - Egyptian Network of High Energy Physics, ASRT-ENHEP, Cairo, Egypt}
\address[add9]{Helwan University \& CTP, Cairo, Egypt}
% ...
\address[add14]{Texas A\&M University at Qatar, Doha, Qatar}
% ...
\address[add16]{University of Florida, Gainesville, U.S.A.}
% ...
\address[add18]{CERN, Geneva, Switzerland}
% ...
\address[add20]{National Center for Physics, Quaid-i-Azam University Campus, Islamabad, Pakistan}
% ...

\begin{abstract}
A novel approach which uses Fibre Bragg Grating (FBG) sensors has been
utilised to assess and monitor the flatness of  Gaseous Electron
Multipliers (GEM) foils. The setup layout and preliminary results
are presented.

\end{abstract}

\begin{keyword}
FBG sensors \sep Triple-GEM detector \sep Mechanical stretching \sep foils planarity \sep gas detectors

    \PACS 29.40.Cs \sep 29.40.Gx    

\end{keyword}

\end{frontmatter}

\section{FBG sensors as a strain measurement}

\begin{figure}[htb]
  \centering
  \includegraphics[width=0.45\textwidth]{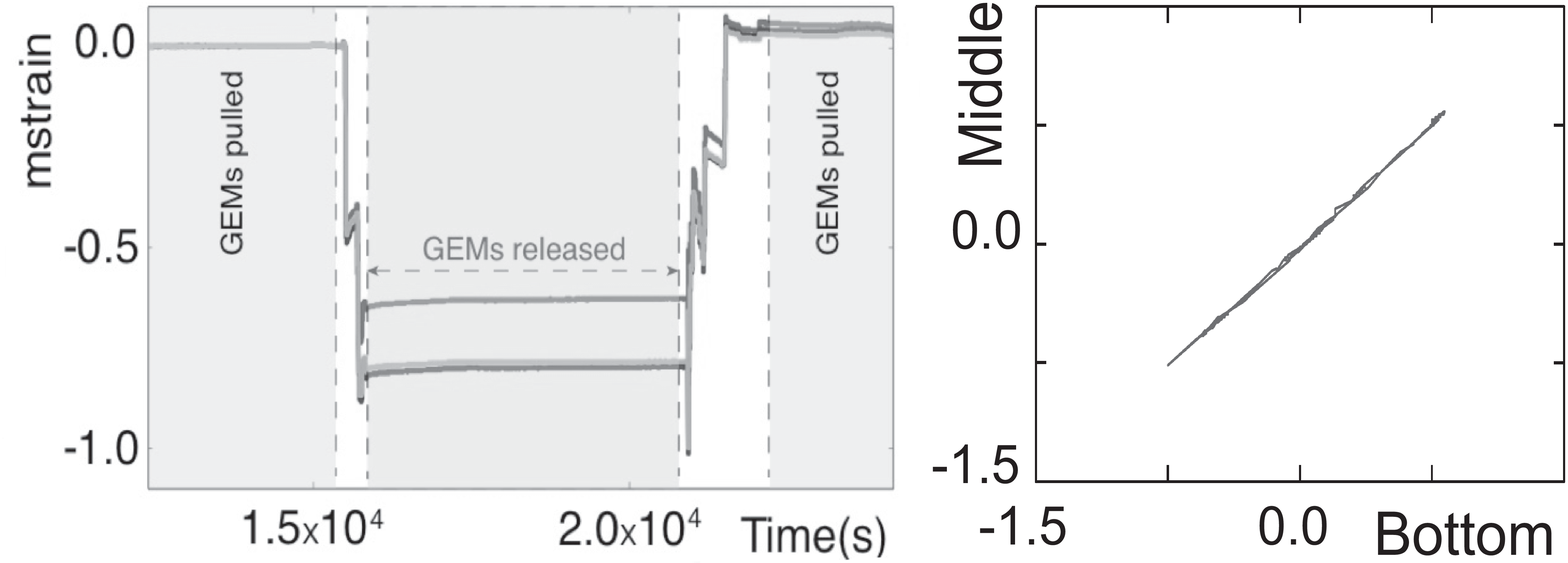}
  \caption{(Left)Three regions corresponding to the mechanical
    stretched, loose and again stretched triple GEM foils stack
    respectively. (Right) The correletion of the strains measured in
    two different foils of the stack}
  \label{fig:subfig2}
\end{figure}

To upgrade the Compact Muon Solenoid (CMS) muon system 144 GEM
chambers will be installed in the high pseudorapidity region of CMS
during Long Shutdown 2 (LS2) of the Large Hadron Collider
\cite{Abbaneo:2014zxc}. The GEMs can provide extra leverage on
precision studies of standard model physics, as well as open up a
window to explore exotic signatures with muons in the high eta region
\cite{Abbaneo:2014lja}.  The GEM chambers will
be located close to the beam pipe where a high flux of low Pt muons is
expected.  The GEM chambers can easily handle this rate due to their
high rate capability of 100 MHz/cm$^2$.  The large active area of each
GE1/1 (GEM Endcap) chamber, approximately 0.4 m$^2$
\cite{Colaleo:2021453}, consists of a triple-GEM foil stack. These
foils need to be stretched simultaneously in order to secure the
planarity and consequent uniform performance of the GE1/1 chamber
\cite{Abbaneo:2013nba}. The GE1/1 detector technology used for CMS is
described in detail in these same conference proceedings (Elba 2015)
by Gilles De Lentdecker with title "Status Report of the Upgrade of
the CMS muon system with triple-GEM detectors". 
The FBG sensors act as low cost precision spatial and temperature sensing tools and
they are commonly used for strain measurements \cite{Benussi:2010yw}
\cite{Caponero:2012py} \cite{Benussi2012483}. In this work
FBG sensors are used to measure the planarity and mechanical tension
of the GEM foils in the GE1/1 chambers. 
A FBG is a type of distributed Bragg reflector, constructed in a short
segment of optical fiber that reflects particular wavelengths of light
and transmits all others. The sensitivity of FBG in terms of strain,
defined as relative elongation w.r.t. the initial position is of the
order of 0.1 micron. This is achieved by creating a periodic variation
in the refractive index of the fiber core, which generates a
wavelength-specific dielectric mirror. Therefore it can be used as a
strain measurement tool since variation of the FBG translates into
different light frequency response.
In order to validate the mechanical stretching technique a network of
FBG sensors is affixed on the triple-GEM stack.
Each sensor is glued on the GEM foil using a very thin layer of
epoxy glue. The test
is performed by modifying the stretching conditions of the GEM foils
stack with real time monitoring and recording of the FBG sensors data.
The test starts with the chamber normally assembled with the GEM stack
mechanically stretched to the nominal tensile load.  After some time
while steady in the starting condition, the mechanical stretching of
the GEMs is released and kept in such condition for several hours.
Finally the GEMs are stretched again up to the nominal tensile
load. The trends of the FBG sensors are shown in figure \ref{fig:subfig2}(Left).
The steep variations of the strain evident in figure \ref{fig:subfig2}(Left)
correspond to the actions of un-screwing and screwing the mechanical stretchers
during the test. The initial stretch value is assumed as reference
condition with strain = 0. When stretchers are un-screwed the strain
goes to the lower value, different strain values apply to different
foils as they fold quasi-free and assume unequal conditions. After the
stretchers are screwed back, the strain value is similar for all
foils, showing that they all experience similar stretching, about the
original value of the reference condition. Thus it can be inferred
that at the predetermined tensile load all foils reach a similar
stretched level although they started from different values. From the
plot it can be seen that all the sensors of the network react at the
same moment. These results allow us to validate the mechanical
stretching assembly technique for GE1/1 chambers. Further tests are
ongoing to confirm other important parameters such as the optimal
tensile load to be applied to the GEMs and the maximum planarity
obtainable for the GEMs without applying a load beyond the Young's
region for GEM foils.

In figure \ref{fig:subfig2}(Right), the mutual comparison of two GEM foils
(the bottom and the middle ones) shows the almost perfect correletion
between the two strian measured demonstrating that all the foils
realize almost the same strain during the assembly.
This shows that the adopted stretching technique is
validated at nominal tensile stress.

\section{Conclusion}

By using the FBG sensors we successfully demonstrated that the
mechanical stretching technique adopted to assemble the GE1/1 chambers
is reliable and secures the correct tensioning of the three foils. By
applying the correct tension across the GEM stack a uniform gap
spacing can be obtained, which is extremely
important to get the required performance of the detector. Several
tests are ongoing by using the same FBG sensors to optimize the
tensile load in order to avoid damage and guarantee planarity of the
GEM foils.

\section* {Acknowledgments}
We gratefully acknowledge the support of FRS-FNRS
(Belgium), FWO-Flanders (Belgium), BSF-MES (Bulgaria),
BMBF (Germany), DAE (India), DST (India), INFN (Italy),
NRF (Korea), QNRF (Qatar), and DOE (USA).

\end{document}